\documentclass[aps,amsmath,amssymb,graphicx,longbibliography,groupedaddress]{revtex4-1}
\usepackage{caption}
\usepackage{subcaption}
\usepackage [latin1]{inputenc}
\usepackage{amsmath}
\usepackage{amssymb}
\usepackage{listings}
\usepackage[margin=1.3in]{geometry}
\usepackage{graphicx}
\usepackage{array}

\newcommand{\mean}[1]{\ensuremath{\left< #1 \right>}}
\newcommand{\de}[1]{\textrm{d}#1}

\begin{document}
\pagenumbering{arabic}
\title{A phenomenological spatial model for macro-ecological patterns in species-rich ecosystems}
\author{Fabio Peruzzo \& Sandro Azaele
\vspace{.5cm}}

\affiliation{Department of Applied Mathematics, School of Mathematics, University of Leeds, Leeds LS2 9JT, United Kingdom.
\\ Email: S.Azaele@leeds.ac.uk
\vspace{.5cm}}

\begin{abstract}
Over the last few decades, ecologists have come to appreciate that key ecological patterns, which describe ecological communities at relatively large spatial scales, are not only scale dependent, but also intimately intertwined. The relative abundance of species -- which informs us about the commonness and rarity of species -- changes its shape from small to large spatial scales. The average number of species as a function of area has a steep initial increase, followed by decreasing slopes at large scales. Finally, if we find a species in a given location, it is more likely we find an individual of the same species close-by, rather than farther apart. Such spatial turnover depends on the geographical distribution of species, which often are spatially aggregated. This reverberates  on the abundances as well as the richness of species within a region, but so far it has been difficult to quantify such relationships. 

Within a neutral framework -- which considers all individuals competitively equivalent -- we introduce a spatial stochastic model, which phenomenologically accounts for birth, death, immigration and local dispersal of individuals. We calculate the pair correlation function -- which encapsulates spatial turnover -- and the conditional probability to find a species with a certain population within a given circular area. Also, we calculate the macro-ecological patterns, which we have referred to above, and compare the analytical formul\ae \ with the numerical integration of the model. Finally, we contrast the model predictions with the empirical data for two lowland tropical forest inventories, showing always a good agreement.

\end{abstract}

\maketitle
\section{Introduction}
In recent years important contributions to our understanding of community assembly and spatial ecology have come from the study of ecological patterns across scales \cite{rosindell2007species,McGill2007,storch2012universal,grilli2012spatial,Azaele2016}. Macroecology has been prolific at suggesting a wealth of interesting patterns and mechanisms \cite{brown1995}. 

For instance, considerable effort has been spent in understanding patterns such as the Relative Species Abundance (RSA) -- which gives the probability of finding a species with $ n $ individuals living on a specific area. The RSA has a pivotal role in identifying the drivers of commonness and rarity in species-rich ecosystems, including tropical forests and coral reefs \cite{Volkov2003,Volkov2005,Volkov2007,Azaele2015}, and has multi-faceted implications, including conservation strategies. This has stimulated a number of approaches attempting to explain the mechanisms underpinning the RSA curve, and there is an ongoing debate over the relative superiority of the proposed models without producing, however, a conclusive answer \cite{McGill2007,may2016abundance}. So far, one of the main issues has been that many reasonable models are able to match empirical data fairly well, thereby hampering the possibility to support a particular theory. This suggests that we should prefer a model over another one, depending on its ability to produce multiple predictions -- in addition to the original pattern -- in agreement with empirical data and without any further parameter fitting. In many cases, authors have tried to explain empirical RSAs by means of stochastic, mean-field models which assume well-mixed populations \cite{Azaele2006,Black2007,Etienne2007b}, which usually are not. 

In contrast, spatial aspects of biodiversity have been described by the so-called $ \beta $-diversity, which overtakes the assumption of individuals placed uniformly at random in space by capturing key aspects of the spatial distribution of species, such as their characteristic spatial turnover. Indeed, classical approaches to population ecology have commonly overlooked the empirical finding that real populations are spatially clustered across a wide range of scales. However, spatial aggregation is important because it increases the turnover of species in space and therefore decreases the similarity of communities that are farther apart \cite{plotkin2000species,azaele2009predicting}. One of the simplest ways to capture this similarity decay with spatial separation is to introduce the Pair Correlation Function (PCF) \cite{Azaele2015,morlon2008general}, which can be defined -- as we will do in the following -- as the correlation in species' abundances of a pair of samples at a given distance.

Finally, another empirical pattern that has received a remarkable attention and has a long history of research is the Species-Area Relationship (SAR) \cite{storch2012universal,grilli2012spatial,Azaele2015,Arrhenius1921} -- which describes how the average number of species increases with the size of the sampled area. This is considered one of the most important and, probably, universal ecological patterns, although the understanding of the underlying mechanistic causes of the SAR curve have progressed slowly and only recently. 

The macroecological patterns that we have described so far are not independent from one another. Theoretical ecologists have been developing an understanding of the relationships among these patterns, and there is a growing appreciation that such macro-ecological measures of biodiversity are inter-related in a deep way. Since Harte and colleagues \cite{harte1999estimating} first suggested that it should be possible to estimate the SAR for a region by examining scattered point survey data, several models have emerged. Some of them are purely geometrical \cite{vsizling2004power} or based on the application of the maximum entropy to ecology \cite{Harte2011}; other studies have also reported the effects of particular biological traits on the shape of SAR \cite{drakare2006imprint}. Here, for the sake of simplicity and to make analytical progress, we will assume neutral population dynamics \cite{Azaele2016,Hubbell2001,Alonso2004,Rosindell2010,Rosindell2011}.

The neutral theory of biodiversity is a theoretical framework for ecological communities with one trophic level, i.e. for species which compete for the same pool of limited resources. Examples are plant species in a forest, breeding birds in a large geographical region, hoverflies living in certain landscapes or coral colonies thriving in warm and shallow waters reachable by sunlight. In the neutral approximation all individuals have the same chances to die or survive and their competition does not depend on the species they belong to. Besides, the population dynamics is assumed to be fundamentally stochastic. Therefore, from the neutral standpoint, individuals' stochastic dynamics is more important than species identity, when it comes to explaining empirical community patterns. However crude and unrealistic these assumptions may look like, they are at the core of models that are in good agreement with empirical measurements at stationarity. Despite such agreements do not necessarily imply that the population dynamics is neutral at the individual level, neutral theory is useful to unveil universal community patterns and it is, probably, more valuable when it fails than when it matches the data. Falsifying one or more of its assumptions, in fact, may inform key aspects of community dynamics.
 
In the following we will focus on a phenomenological neutral model, whose dynamics is spatially-explicit and stochastic. Because it cannot be solved explicitly in full generality, we will introduce a method for calculating analytically approximate formul\ae \ for the three patterns which we have alluded to above. We will then compare the analytical expressions with the numerical integration and, finally, we will show that the model is able to describe the empirical RSA, SAR and PCF of two tropical forests which harbour hundreds of plant species. With this model one can translate information from one pattern to another and extrapolate patterns outside the region of parametrization. \\

\subsection{The RSA in the mean-field approximation}
Before introducing the spatial stochastic model, in order to make it clear how the neutral assumption enters the definition of a model, let us first focus on a simple form of RSA that can be deduced at the mean-field level. If we assume that the dynamics is Markovian and described by a birth and death (one-step) process, then in general the birth and death rates of species $ \alpha $ can be written down as $ b_{\alpha}(n_1,n_2,\ldots,n_S) $ and $ d_{\alpha}(n_1,n_2,\ldots,n_S) $, respectively, where $ n_i $ is the population size of species $ i $ and $ S $ is the total number of species in a given region. If interactions are neutral, then those rates should be symmetric functions of species' population sizes and should not depend on the species label $ \alpha $ (strictly speaking, this defines a symmetric model \cite{Azaele2016} -- not a neutral one --, but in the following we will not make such a distinction). Also, if we further assume that species are independent, then the birth and death rates factorise and we can focus on the dynamics of just one species, because any species is not affected by the presence of the others. In this way, the neutral and the independence assumptions allow us to think of the population sizes of species as independent realizations of a stochastic process. In our case, the birth and death rates are $b_n$ and $d_n$, respectively, with $ n $ the number of individuals of a species in a given region. Therefore, the time evolution of the probability distribution of $ n $ is described by the following master equation
\begin{equation}
\frac{\partial p_n(t)}{\partial t}=p_{n+1}(t) \ d_{n+1}+p_{n-1}(t) \ b_{n-1}-p_n(t) \ (b_n+d_n)\quad,
\label{eq:ME}
\end{equation}
where $p_n(t)$ is the probability that a species has $n$ individuals at time $t$. Of course, this equation needs to be equipped with boundary conditions that prevent $ n $ from becoming negative. Common choices are either reflecting or absorbing boundaries, depending on the nature of the problem. When $ n=0 $ is reflecting, the equilibrium solution can be easily calculated \cite{VanKampen1992} and is, for $n>0$,
\begin{equation}
P_n=P_0 \prod_{i=0}^{n-1} \frac{b_i}{d_{i+1}} \quad,
\label{eq:statiobase}
\end{equation}
where $P_0$ is a normalization constant. If individuals belonging to abundant and rare species have the same chances to die, or survive and give birth to an offspring, then the per capita birth and death rates cannot depend on $ n $ and therefore, for $ n\geq 0 $, we have to set
\[
b_n=g n +\delta_{n,0}\nu \qquad d_n=r n\quad,
\]
where $g$ and $r$ are positive constants, and $\nu$ is the speciation rate. In this framework there is no explicit biological mechanism for speciation: $ \nu $ is a parameter that ensures that the system is always populated by one individual whenever species go extinct (reflecting boundaries). Let's denote by  $\Phi_n$ the number of species with $n$ of individuals. If $S$ is the empirical number of species in our ecosystem, from eq.\eqref{eq:statiobase} we get
\begin{equation}
\langle \Phi_n \rangle=S P_0 \prod_{i=0}^{n-1} \frac{b_i}{d_{i+1}}=S P_0 \frac{b_0 b_1 ... b_{n-1}}{d_1 d_2 ... d_n}=\theta \frac{x^n}{n}\quad,
\label{eq:fisherlog}
\end{equation}
where $x=g/r<1$, $ n>0 $ and $\theta=S P_0 \nu/g$ is known as the biodiversity parameter. Eq.(\ref{eq:fisherlog}) is known amongst ecologists as `Fisher log-series', and was first discovered experimentally in 1943 \cite{Fisher1943}. This distribution has no internal mode and therefore it predicts that singleton species (i.e., those with one individual only) are always the most frequent. This is not always the case, as many communities have species' abundances that are more frequent than singletons. These RSAs can be more adequately explained with an alternative choice of rates, i.e.
\begin{equation}
b_n=g n + b \qquad d_n=r n\quad,
\label{ratesNB}
\end{equation}
where the parameter $ b >0$ incorporates immigration. Ultimately, in this setting rare species have a mild reproductive advantage over the more common ones. The equilibrium solution is the following negative binomial distribution
\begin{equation}
\langle \Phi_n \rangle=S (1-x)^{\frac{b}{g}}\left(\frac{b}{g}\right)_{\!\!n} \frac{x^n}{n!}\quad,
\label{eq:negbin}
\end{equation}
where $ (a)_n=a(a+1)\ldots (a+n-1) $  with $ (a)_0=1 $, $ n=0,1,\ldots $ and $ x=g/r $ with $ 0<x<1 $. This distribution can produce an internal mode in species' abundances and predicts that communities should harbour only a few species that are common and many species that are rare. This RSA is more flexible than the Fisher log-series and is in good agreement with empirical data \cite{Azaele2006,Volkov2007}.

\subsection{A mean-field  Langevin equation for the RSA}
\label{sec:meanfield}
Larger areas of species-rich communities often sustain larger populations and support more species because, typically, they encompass greater habitat diversity and richer pool of resources. This simple observation shows that community patterns at relatively large spatial scales might be described by models which treat population size as a continuous random variable. Also, it suggests to include the principal effects driving the macro-ecological patterns in a simplified, phenomenological fashion. Within the neutral approach and assuming that the effects we outlined in the previous section are the most important driving factors, we get the following Fokker-Planck (FP) equation for the diffusive approximation of the master equation (eq.(\ref{eq:ME})) with rates defined in eqs.(\ref{ratesNB})
\begin{equation}
\frac{\partial P(n,t)}{\partial t}=-\frac{\partial}{\partial n} \Bigl[ (b-\mu n) \ P(n,t) \Bigr]+\sigma^2\frac{\partial^2}{\partial n^2} \Bigl[ (n+\epsilon) \ P(n,t)\Bigr]\quad,
\label{eq:FP}
\end{equation}
where $\mu=r-g>0$, $\sigma^2=(r+g)/2$ and $ \epsilon=b/(r+g) >0$. The equilibrium solution of this equation provides the continuous RSA, i.e.

\begin{equation}
P(n)=P_0 (n+\epsilon)^{\frac{b+\mu  \epsilon}{\sigma ^2}-1}e^{-\frac{\mu  n}{\sigma ^2}}\quad,
\label{eq:statio1}
\end{equation}
where $ P_0 $ is a normalization constant. A given large region is usually affected by a small immigration rate, which hence suggests that $ \epsilon $ is typically a small parameter. If we treat it as such, then eq.(\ref{eq:statio1}) can be approximated (at zeroth order) by a (normalized) gamma distribution of the form:
\begin{equation}
P(n)=\Bigl( \frac{\mu}{\sigma^2} \Bigr)^{\frac{b}{\sigma^2}} \ \frac{n^{\frac{b}{\sigma^2}-1} e^{-\frac{\mu n}{\sigma^2}}}{\Gamma( b/\sigma^2)}\quad,
\label{eq:statio}
\end{equation}
where $ \Gamma(x) $ is the gamma function. The (non-uniform) correction to this equation is of order $ \epsilon\ln \epsilon $ for $ b/\sigma^2\geq 1 $ and $ \epsilon^{b/\sigma^2} $ for $ 0<b/\sigma^2<1 $. In real species-rich ecological communities one typically finds $ r\simeq g $ (usually, $1-g/r<0.02$, hence $ \mu$ is positive and small \cite{Volkov2005}), which therefore allows the existence of a few species with a large number of individuals (population sizes larger than $ \sigma^2/\mu=r/(r-g) $ when $ 0<b/\sigma^2<1 $ and $ \epsilon\ll 1 $). Rare species, instead, have population sizes typically smaller than $ b/\mu=b/(r-g) $ (for $ 0<b/\sigma^2<1 $ and $ \epsilon\ll 1 $). As expected, eq.(\ref{eq:statio}) is the equilibrium solution of the simpler FP equation
\begin{equation}
\frac{\partial P(n,t)}{\partial t}=-\frac{\partial}{\partial n} \Bigl[ (b-\mu n) \ P(n,t) \Bigr]+\sigma^2\frac{\partial^2}{\partial n^2} \Bigl[ n P(n,t)\Bigr]\quad,
\label{eq:FPnb}
\end{equation}
which corresponds to the Langevin equation (in the It\={o} prescription)
\begin{equation}
\dot{n}=b-\mu n +\sigma \sqrt{n} \ \xi(t)\quad,
\label{eq:langevin}
\end{equation}
where $\xi(t)$ is a zero mean white noise with $\langle \xi(t) \xi(t') \rangle= 2\delta(t-t')$. Eq.(\ref{eq:langevin}) has a nice interpretation: in the limit of a small immigration rate, the dynamics of the RSA results from the trade off between net immigration and net death rates (i.e., $ b-\mu n $), and the fluctuations about these deterministic terms are simply driven by the central limit theorem (i.e., fluctuations $\propto \sqrt{n}  $). The agreement of eq.(\ref{eq:statio}) with the data \cite{Azaele2006}, therefore, suggests that demographic stochasticity may play a major role in sculpting macroecological patterns, including the RSA. In the following section these considerations will form the backbone of the spatial version of the model, thus extending the importance of the effects of immigration, birth, death and demographic stochasticity to spatial patterns as well.  

\section{A phenomenological spatial stochastic model: linking macro-ecological patterns}
The assumption of well-mixed populations, of course, cannot account for the spatial turnover of species and the increase of species richness with sampled area. These two patterns are captured by the PCF and SAR, respectively, as explained in the introduction. A region with a high rate of spatial turnover of species, in which the PCF decays steeply, has also a steep increase in the SAR, because a given area contains relatively more species compared to other regions where the PCF decays more gradually. Also, empirical data highlight that the PCF is, typically, a monotonically decreasing function of distance. This underlines the important role of spatial clumping of individuals, because were an individual found somewhere, it would be more likely to find another one close by.

These observations lead naturally to a simple spatial extension of the continuous model of the RSA. Since we are interested in spatial patterns at relatively large scales, we consider a phenomenological generalization in which space is coarse grained. We assume space is partitioned by a mesh into a collection of voxels -- or, more precisely, a regular graph (or lattice) in which each vertex has 2$ d $ nearest neighbours, being $ d $ space dimension. Within each voxel (or, equivalently, vertex or site, which hereinafter will be used as synonyms), individuals are considered well-mixed, diluted and treated as point-like particles which undergo the demographic dynamics defined by eq.(\ref{eq:langevin}), which incorporates birth, death and immigration (in the language of chemical reaction kinetics, these are first-order reactions known as autocatalitic production, degradation and production from source, respectively). 

As the customary approach in the reaction-diffusion master equation (RDME), we will assume that, within a hypercubic voxel of width $ a $ ($ a $ is the lattice spacing as well), individuals are uniformly placed at random in space (i.e., voxels have no internal spatial structure). Therefore, $ a $ should be much smaller than all the other macroscopic length scales of interest, including the characteristic spatial correlation length of the system. In the following numerical integration and empirical analysis, this will always be the case. 
The set of coupled stochastic differential equations defining the model are

\begin{equation}
\dot{n}_i(t)=D \nabla^2_i n_i(t)+b-\mu n_i(t) +\sigma \sqrt{n_i(t)} \ \xi_i (t)\quad,
\label{eq:SPDE}
\end{equation}
where $n_i(t)$ is the density of individuals in the $i$-th site at time $ t $, $\xi_i(t)$ is a zero mean white noise (depending on site $i$) with correlation $\langle \xi_i(t) \xi_j(t') \rangle=2\delta(t-t') \delta_{i,j}$. $ D $ is the ``diffusion'' coefficient and 
\begin{equation}
\nabla^2_i n_i(t) =  \frac{1}{a^2}\sum_{j\in \partial(i)}[n_j(t)-n_i(t)]\quad,
\label{def:discreteNabla}
\end{equation} 
where $ \partial(i) $ indicates the set of nearest neighbours of $ i $. There is nothing special about our choice of local movement, more general connectivities could have been chosen to study the effects of different topologies on macroscopic patterns \cite{muneepeerakul2011evolution}. More importantly -- and unlike the RDME approach --, here individuals move locally on the mesh in a deterministic fashion, as governed by the discrete Laplacian. This is tantamount to neglect contributions to stochasticity due to the random hopping of individuals, which is expected to be a good approximation for large diffusion constants \cite{dean1996langevin}. Therefore, linear reactions taking place inside voxels -- independent of diffusion -- are supposed to be the main source of stochasticity in the system. In this framework, individuals do not undergo a continuous time random walk on the mesh, as can be seen from eq.(\ref{eq:SPDE}) when the internal demographic dynamics is switched off. This is one of the main reasons why this spatial stochastic model, at least in the current formulation, cannot be considered an appropriate coarse-grained approximation of an underlying microscopic, spatially continuous model. However, these approximations are not expected to have large effects on the first two moments, which we will study in the following sections and are at the core of our analysis. This is only a phenomenological framework which provides an analytical way to calculate macroecological patterns, starting from simple yet important demographic and spatial factors. Yet, microscopic models which are continuous in space, such as independent branching Brownian processes (or superprocesses \cite{etheridge2000introduction}), might probably have a discrete approximation close to the current formulation. This will be investigated in a future work.

If we indicate with $ \{ n \}$ a given configuration of population sizes on the lattice, i.e., $ \{ n \}=\{ n_1, n_2, ...\} $, the probability density function of $ \{ n \}$, $P(\{ n \})$, satisfies the following FP equation (\textit{sensu} It\={o})
\begin{equation}
\partial_t \ P ( \{ n \},t )=-\sum_z \frac{\partial}{\partial n_z} \Bigl[ \Bigl( D \nabla^2_z n_z(t) +b-\mu n_z \Bigr) P ( \{ n \},t ) \Bigr]+\sigma^2 \sum_z \frac{\partial^2}{\partial n_z^2} \Bigl[ n_z  P(\{ n\},t)\Bigr]\quad,
\label{eq:FPtotal}
\end{equation}
where the sums are over all sites of the lattice. It is easy to see that the average density per site is $ \langle n_i \rangle=b/\mu $. It is interesting to notice that this model has a non-trivial stationary distribution only for $ b>0 $ and when the per capita death rate is strictly larger than the per capita birth rate (i.e., $ \mu>0 $), because of the lack of a carrying capacity. In this sense, it is a minimal model for calculating large scale patterns: if one sets to zero one or more parameters, then the predicted macro patterns -- if they exist -- are trivial.

\subsection{Calculating the Pair Correlation Function}
The PCF describes the correlation in species' population abundances between different spatial locations. As we mentioned before, it plays a crucial role in linking some of the most important macroecological patterns. 

Let's consider two sites $i$ and $j$ in a ($ d $-dim) lattice and calculate $\langle n_i  n_{j} \rangle  $. Multiplying eq.(\ref{eq:FPtotal}) by $ n_i n_j $ and integrating all $ n $'s from zero to infinity (or using the usual  It\={o} formula with eq.(\ref{eq:SPDE})), one finds the equation for the time evolution of  $\langle n_i  n_{j} \rangle $, i.e.,

\begin{equation}
\frac{\partial}{\partial t} \mean{n_i n_j}= D(\nabla^2_i \mean{ n_i n_j} + \nabla^2_j\mean{ n_i  n_j})+ 2b\mean{n} - 2\mu \mean{n_i n_j} +  2\sigma^2\mean{n}\delta_{ij}\quad,
\label{eq:PCFcalc5}
\end{equation}
where $ \mean{n}=b/\mu $ and $ \delta_{ij} $ is the Kronecker delta. Because we are interested in stationary patterns, we drop the time derivative and simplify the equation by looking at the correlation $ G_{i,j}=\mean{n_i n_j}-\mean{n_i}\mean{n_j} = \mean{n_i n_j}-\mean{n}^2$. $G_{i,j}$ actually satisfies

\begin{equation}
D( \nabla^2_i G_{i,j}+ \nabla^2_jG_{i,j}) - 2\mu G_{i,j} +  2\sigma^2\mean{n}\delta_{ij}=0\quad.
\label{eq:pcfG}
\end{equation}
In order to solve this equation, let us introduce a system of Cartesian coordinates and indicate with $ \textbf{x} $ the $ d $-dim position vector of a site. Basically, in the previous equation we make the substitution $ i \rightarrow \textbf{x} $ and $ j \rightarrow \textbf{y} $, with the agreement that changes in any direction in the coordinates have to be made in multiples of $ a $, the lattice spacing. In this way,  we can use Fourier series to find an expression for $ G_{\textbf{x},\textbf{y}} $ in an infinite lattice. After some algebraic manipulations, we finally get
 
 \begin{equation}
 G_{\textbf{x},\textbf{y}} = \left(\frac{a}{2\pi}\right)^d \frac{\sigma^2 b}{\mu^2} \int_{\mathcal{C}}\de{\textbf{p}}\ \frac{e^{i \textbf{p}\cdot (\textbf{x}-\textbf{y})}}{1+\frac{2D}{\mu a^2}\sum_{i=1}^{d}(1-\cos(p_i a))}\quad,
 \label{eq:solDiscrete}
 \end{equation}
where $ p_i $ is the $ i $-th Cartesian component of $ \textbf{p} $ and $\mathcal{C}  $ is the hypercubic ($ d $-dim) primitive unit cell with size $ 2\pi/a $. As expected, $ G_{\textbf{x},\textbf{y}} $ is translational invariant and in $ d=1 $ reduces to a simple exponential:

 \begin{equation}
 G_{x,y} = C k^{|x-y|/a}\quad,
 \label{eq:solDiscreteExp}
 \end{equation}
where $ x, y= 0, a, 2a, \ldots $; $ k<1 $ and $ C $ are positive constants which can be either calculated from eq.(\ref{eq:solDiscrete}) or by direct substitution into eq.(\ref{eq:pcfG}). For $ k $ one gets
\begin{equation}
k=1+\frac{\mu a^2}{2D} - \sqrt{\frac{\mu^2 a^4}{4D^2}+\frac{\mu a^2}{D}}\quad,
\label{sol:zeta}
\end{equation}
from which one deduces the correlation length $\xi=-a/\ln(k) $. Notice that $ \xi \rightarrow \sqrt{D/\mu} $ when $ a \rightarrow 0 $.

Instead of trying to calculate explicitly the integral in eq.(\ref{eq:solDiscrete}), we can obtain a good deal of simplification and insight by taking its continuum spatial limit (i.e., $ a\rightarrow 0 $ and the parameters are appropriately re-defined). Such a limit leads to 

\begin{eqnarray}
\mathcal{G}(\textbf{x},\textbf{y}) &=&\frac{1}{\left(2\pi\right)^d} \frac{\sigma^2 b}{\mu^2} \int_{{\mathbb{R}}^d}\de{\textbf{p}}\ \frac{e^{i \textbf{p}\cdot (\textbf{x}-\textbf{y})}}{1+\frac{D}{\mu}\textbf{p}^2}\notag\\
&=& \frac{\hat{\rho}^2\mean{n}^2}{\left(2\pi\hat{\lambda}^2\right)^{d/2}} \left(\frac{|\textbf{x}-\textbf{y}|}{\hat{\lambda}}\right)^{(2-d)/2}K_{(2-d)/2}\left(\frac{|\textbf{x}-\textbf{y}|}{\hat{\lambda}}\right),
 \label{eq:solContinuous}
\end{eqnarray}
where $ K_{\nu}(x) $ is the modified Bessel function of the second kind of order $ \nu $ or Macdonald's function \cite{Lebedev}, $\textbf{x}$ and $\textbf{y} $ are now continuous vector coordinates and 

\[
\hat{\lambda}=\sqrt{\frac{D}{\mu}} \quad, \quad \hat{\rho}=\sqrt{\frac{\sigma^2}{b}}\quad
\]
are constants with length dimension when $ d=2 $. As expected,  $\mathcal{G}(\textbf{x},\textbf{y}) $ is also the solution of the continuum spatial limit of eq.(\ref{eq:pcfG}) in Cartesian coordinates (and dimension $ d $), i.e.

\begin{equation}
D\nabla^2_{\textbf{z}}\mathcal{G}(\textbf{z})-\mu \mathcal{G}(\textbf{z})+ \sigma^2\mean{n}\delta(\textbf{z})=0\quad,
\label{eq:cont_lim_G}
\end{equation}
where $ \textbf{z}= \textbf{x}-\textbf{y}$, $ \delta(\textbf{z}) $ is a Dirac delta and we took advantage of the translational symmetry of the system.

Of course, $\mathcal{G}$ obtained in eq.(\ref{eq:solContinuous}) may be a good approximation of the discrete correlation only for $|\textbf{x}-\textbf{y}|\gg a  $. As a first approximation, however, one may introduce a lower cut-off to $ \mathcal{G} $ by stipulating that $ \mathcal{G}(\textbf{z})= G_{\textbf{x},\textbf{x}}$ for all $ |\textbf{z}|\leq a $. Because  $ K_{\nu}(x) $ decays exponentially fast for large $ x $ \cite{Lebedev}, eq.(\ref{eq:solContinuous}) also suggests that $ \hat{\lambda} $ is the spatial correlation length of the system. Therefore, this continuous framework works under the condition that $  \hat{\lambda}\gg a $, which is always satisfied in the following analysis.

In the next sections we look into the stationary Pair Correlation Function (PCF) defined as 
\begin{equation}
g_{\textbf{x},\textbf{y}}=\frac{\mean{n_{\textbf{x}}n_{\textbf{y}}}}{\mean{n}^2} \quad, 
\label{def:PCF}
\end{equation}
because -- in a first approximation -- it allows one to study the empirical properties of $ g_{\textbf{x},\textbf{y}} $ independently of $ a $, the spatial resolution introduced to calculate the PCF from the data. As an analytic expression, we will use its continuous version, i.e. $ g(\textbf{x},\textbf{y})= 1+\mathcal{G}(\textbf{x},\textbf{y})/\mean{n}^2$, where $ \mathcal{G}(\textbf{x},\textbf{y}) $ is given in eq.(\ref{eq:solContinuous}) with $ d=2 $. Hence, the PCF reduces to
\begin{equation}
g(r)=1+\frac{1}{2 \pi} \Bigl( \frac{\hat{\rho}}{\hat{\lambda}}\Bigr)^2 K_0 \Bigl( \frac{r}{\hat{\lambda}} \Bigr)\quad,
\label{eq:PCF}
\end{equation}
where $ r=|\textbf{x}-\textbf{y}| $. We will always assume that $ r $ is much larger than $ a $.

\section{A method for calculating macroecological patterns}
The model defined in eq.(\ref{eq:SPDE}) is linear and therefore all the stationary $ n $-point correlation functions can be calculated explicitly. However, having all correlation functions is not sufficient, in general, to build up a closed-form solution of the model, from which one derives all interesting patterns. 

The spatial Relative Species Abundance (sRSA) is defined as the probability that a species has $ n $ individuals within a certain area $ A $, if there are $ S_0 $ species in total in the larger area $ A_0 $ where $ A $ is contained. Therefore, the sRSA is given by the conditional probability $ p(n|A,\{S_0,A_0\}) $, and all correlation functions contribute to such distribution in a non trivial way. So, instead of trying to calculate the sRSA from the correlation functions or the generating functional, we introduce an approximation which allows to make some analytical progress. Afterwards, we will check with the numerical integration of the model that such approximations are good, at least in the region of the parameter space which is relevant to the empirical patterns.

Because the calculations turn out to be easier in the continuum space, in what follows we will essentially work with eqs.(\ref{eq:solContinuous}-\ref{eq:cont_lim_G}), bearing in mind that the results in such limit have to be used \textit{cum grano salis}. For simplicity then, let us focus on a circular region, $C$, of radius R and define the random variable 
\begin{equation}
N(R)= \int_C n({\textbf{x}}) \de{\textbf{x}}\quad,
\label{def:NR}
\end{equation}
which gives the number of individuals of a species living on $ C $ at stationarity. Of course, $ \mean{N(R)}=\mean{n}\pi R^2 $, but we can also calculate the variance, $ \textrm{Var}(N(R)) $. From eq.(\ref{eq:solContinuous}) we get
\begin{equation}
\int_C \int_C \mathcal{G}(\textbf{x},\textbf{y}) \de{\textbf{x}}\de{\textbf{y}} = \mean{N(R)^2}-\mean{N(R)}^2 = \textrm{Var}(N(R))
\end{equation}
and the final expression in $ d=2 $ is
\begin{equation}
\textrm{Var}(N(R))=\langle n \rangle \hat{\rho}^2 \langle N(R) \rangle \Bigl( 1-\frac{2 \hat{\lambda}}{R} \frac{K_1(R/\hat{\lambda}) I_1(R/\hat{\lambda})}{K_0(R/\hat{\lambda}) I_1(R/\hat{\lambda})+K_1(R/\hat{\lambda}) I_0(R/\hat{\lambda})} \Bigr)\quad,
\label{eq:sigma}
\end{equation}
where $ I_{\nu}(x), K_{\nu}(x)$ are modified Bessel functions of the first and second kind of order $ \nu $, respectively \cite{Lebedev}. Of course, this formula is reliable only when $ R\gg a $, but it is interesting to notice that for $ R\gg \hat{\lambda} $ the variance to mean ratio tends to a constant, i.e. 

\begin{equation}
\frac{\textrm{Var}(N(R))}{\mean{N(R)}}\simeq \langle n \rangle \hat{\rho}^2 = \frac{\sigma^2}{\mu}\quad,
\end{equation}
which is exactly the ratio one obtains from the mean field model, i.e. eq.(\ref{eq:statio}). Therefore, at stationarity the system reaches non-Poissonian fluctuations and on large spatial scales it is homogenized by diffusion. This is an example of a result that can be proved under quite general conditions \cite{gardiner1984adiabatic}.

As we have alluded to above, a lot of species-rich ecological communities have  per capita birth and death rates that are very close ($1-g/r<0.02$, hence $ \mu$ is positive and small \cite{Volkov2005}). So, we can roughly estimate the variance to mean ratio as
\begin{equation}
\frac{\textrm{Var}(N(R))}{\mean{N(R)}}\simeq \frac{r}{r-g} \gg 1\quad,
\end{equation}
for $ R\gg \hat{\lambda} $. Moreover, when $ r\simeq g $ both the correlation length, $ \hat{\lambda} $, and the correlation time, $ \mu^{-1} $, of the system are very large. This depicts such empirical communities as they were posed close to a critical point, where large fluctuations have a long-time behaviour and are correlated across many spatial scales.

Along the lines we have outlined before, one could in principle write down the expressions for the higher moments of $ N(R) $. However, a deeper insight and more analytical progress can be achieved by introducing the following crucial approximation: we assume that, at stationarity, the random variable $ N(R) $ is distributed according to the probability density function defined in eq.(\ref{eq:statio}) -- the equilibrium solution of the  mean field model -- with appropriate scale-dependent functions, $ \alpha(R) $ and $ \beta(R) $, which we are going to introduce. This is tantamount to assume that the functional form of the sRSA is the same across all spatial scales and hence the dependence on the spatial scale of the sRSA comes only through such functions. We have borrowed this hypothesis from the phenomenological renormalization group \cite{plischke2006equilibrium}.

In order for the gamma distribution in eq.(\ref{eq:statio}) to match the first two moments of $ N(R) $ that we have calculated, we then introduce a shape function, $ \alpha(R) $, and a scale function, $ \beta(R) $, both depending on $ R $. The final approximate sRSA, $ q(N|R) $, has therefore the form
\begin{equation}
q(N|R)=\frac{1}{\beta(R)} \frac{(N/\beta(R))^{\alpha(R)-1}}{\Gamma(\alpha(R))} e^{-N/\beta(R)}\quad,
\label{eq:sSAD}
\end{equation}
where $\Gamma(x)$ is a gamma function. From the properties of the gamma distribution, it is not difficult to show that, if we choose
\begin{equation}
\alpha(R)=\Bigl( \frac{\langle N(R) \rangle}{\sigma(R)} \Bigr)^2 \quad \textrm{and}\quad \beta(R)=\frac{\sigma(R)^2}{\langle N(R) \rangle}\quad,
\label{eq:albet}
\end{equation}
then we match exactly the first two moments, $ \mean{N(R)} $ and $ \textrm{Var}(N(R)) $. We will show that the approximate expression for the sRSA is in good agreement with the numerical integration of the model. With the formula for $ q(N|R) $ one can directly link the sRSA to the PCF. In fact, when fitting the PCF and obtaining $ \mean{n} $ from the data, we can predict the distribution of species' population sizes across all spatial scales by using eq.(\ref{eq:sSAD}). 

Also, since a species can be observed only when it has at least one individual, the probability that a species is present within an area of radius $ R $ is $ \int_1^{\infty} q(N|R)\de{N}$, from which one can calculate the SAR, an important pattern in many applications.

\section{Numerical scheme for the integration of the model}
\label{numerics}
Na\"{\i}ve numerical schemes for integrating eq.(\ref{eq:SPDE}) are affected by severe drawbacks. For instance, if we apply a first-order explicit Euler method to the simpler eq.(\ref{eq:langevin}) (\textit{sensu} It\={o}), we get
\begin{equation}
n(t +\Delta t)=n(t)+\Delta t [b-\mu n(t)]+\sigma \sqrt{\Delta t \ n(t)} N(0,1) \quad,
\label{eq:SDE_euler}
\end{equation}
where N(0,1) is a zero mean normal random variable with variance $1$. It is well known that, starting from $ n(0)>0 $, this method inevitably leads to produce negative values for $ n(t +\Delta t) $, especially when $ n(t) $ is small. Such unphysical densities are even more harmful when integrating stochastic partial differential equations, strongly biasing spatial correlations.

Building on previous methods \cite{pechenik1999interfacial,dornic2005integration}, we introduce a numerical integration scheme which generates (in the weak sense) the field $ n_i(t) $ at stationarity in 2-dim -- the $ d $-dim case is straightforward --, and ensures, by construction, that the density is always non-negative. 

We first write down the discrete Laplacian on a 2-dim lattice of mesh size $ a $, where every site has 4 nearest neighbours. Secondly, we re-write eq.(\ref{eq:SPDE}) as 

\begin{equation}
\dot{n}_{\textbf{x}}(t)= Y_{\textbf{x}}(t)-\Omega n_{\textbf{x}}(t) +\sigma \sqrt{n_{\textbf{x}}(t)} \ \xi_{\textbf{x}} (t)\quad,
\label{eq:SPDE_num}
\end{equation}
where 
\begin{equation}
Y_{\textbf{x}}(t)=\frac{D}{a^2} \sum_{i=1}^4 n_{\textbf{x}+a\textbf{e}_i}(t)+b
\quad \textrm{and} \quad \Omega = \frac{4D}{a^2}+\mu\quad,
\end{equation}
and $\textbf{e}_1=(1,0)$, $\textbf{e}_2=(-1,0)$, $\textbf{e}_3=(0,1)$ and $\textbf{e}_4=(0,-1)$.

The stationary solutions of the FP equations associated to each local Langevin equation for $ n_{\textbf{x}}(t) $, i.e., eq.(\ref{eq:SPDE_num}), are gamma distributions given by eq.(\ref{eq:statio}) in which, locally, $ b \rightarrow Y_{\textbf{x}} $ and $ \mu \rightarrow \Omega $. $ Y_{\textbf{x}} $ is then a new immigration parameter which accounts for the global as well as the local influx of individuals from the 4 nearest neighbours into the site with $ \textbf{x} $ coordinates; $\Omega $ is a new death rate which includes the possibility that individuals leave the site at $ \textbf{x} $ because of diffusion, in addition to the demographic death rate. If we initialize the lattice with $ n_{\textbf{x}}^{(0)}\geq 0 $, we can then update each and every site by sampling from the local gamma distribution, conditioning on the nearest neighbours. Hence, at the $ m+1 $ sampling step the local density is given by

\begin{equation}
n_{\textbf{x}}^{(m+1)}=\textrm{Gamma}\left[\frac{Y_{\textbf{x}}^{(m)}}{\sigma^2},
\frac{\sigma^2}{\Omega}\right]\quad,
\end{equation}
where $ m \in \mathbb{N}$,
\begin{equation}
Y_{\textbf{x}}^{(m)}=\frac{D}{a^2} \sum_{i=1}^4 n_{\textbf{x}+a\textbf{e}_i}^{(m)}+b
\end{equation}
and $ \textrm{Gamma}[\alpha,\beta] $ is a gamma variate with shape parameter $ \alpha $ and scale parameter $ \beta $. One keeps updating the system until all the stationary summary statistics of interest do not change significantly in different generations (or they match a stationary summary statistics calculated analytically from the model).

Because $ n_{\textbf{x}}^{(0)}\geq 0 $ and also $ Y_{\textbf{x}} $ and $ \Omega $ are strictly positive at all steps ($ D, b, \mu $ and $ \sigma $ are all strictly positive), by construction $ n_{\textbf{x}} $ is always non-negative and finite at all steps.

\subsection{Comparisons with analytical solutions}
\begin{figure}[t]
\includegraphics[scale=0.6]{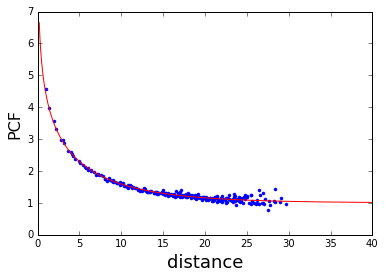}
\caption{Comparison between the analytical (see eq.(\ref{eq:PCF})) and numerical PCF calculated from the stationary densities generated by implementing the numerical scheme outlined in Sec.(\ref{numerics}). Here the parameters are $D=1$, $b=0.005$, $\mu=0.01$, $\sigma=2.1$ and the distance is in lattice spacing units.}
\label{fig:PCF}
\end{figure}

We implemented the numerical scheme on a 200x200 lattice with periodic boundary conditions. Each site was initialized by drawing from a gamma distribution with shape parameter $\alpha=b/\sigma^2$ and scale parameter $\beta=\sigma^2/\mu$. The comparisons between the analytical formul\ae \ obtained in the continuum approximation and the numerical integrations were carried out by considering 1,000 independent realizations at stationarity on the square lattice. The results for the numerical and analytical PCF are shown in fig.\ref{fig:PCF} for the correlation length $ \hat{\lambda}=10 $.

\begin{figure} 
\includegraphics[scale=.8]{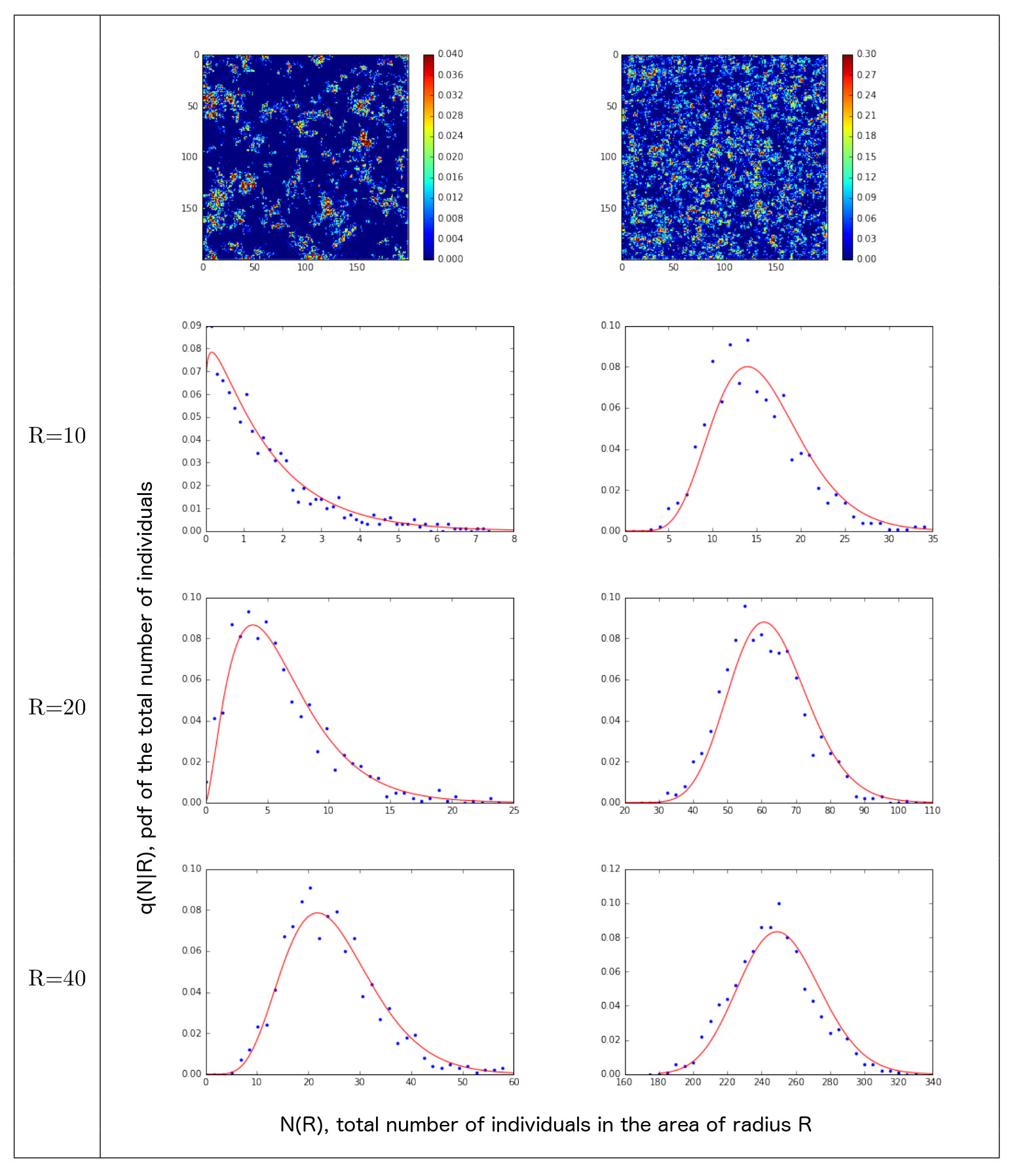}
\caption{Comparison between the analytical (see eqs.(\ref{def:sRSAemp}), (\ref{eq:sSAD}) and (\ref{eq:albet})) and numerical sRSA calculated from the stationary densities generated by implementing the numerical scheme outlined in Sec.(\ref{numerics}). Here the parameters are $D=100$, $b=0.005$, $\mu=1$, $\sigma=2.1$ for the left column and $D=1$, $b=0.005$, $\mu=0.1$, $\sigma=0.5$ for the right column. The two upper panels depict two snapshots of the stationary densities on the corresponding lattices. The radius is in lattice spacing units.}
  \label{fig:SAD}
\end{figure}

\begin{figure} 
\centering
\includegraphics[scale=0.7]{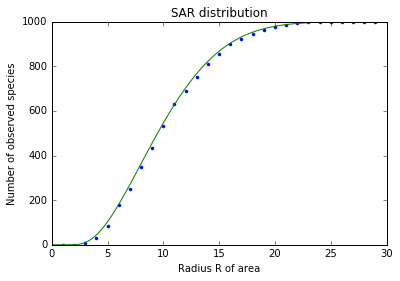}
\caption{Comparison between the analytical (see eqs.(\ref{def:SARemp}), (\ref{eq:sSAD}) and (\ref{eq:albet})) and numerical SAR calculated from the stationary densities generated by implementing the numerical scheme outlined in Sec.(\ref{numerics}). Here the parameters are $D=100$, $b=0.005$, $\mu=1$, $\sigma=2.1$ and the radius is in lattice spacing units.}
\label{fig:SAR}
\end{figure}

For a given realization at stationarity, we decided that a species is observable -- that is, it has at least one individual -- within a given area $C$ of radius $R$, if $N(R)=\sum_{\textbf{x}\in C} n_{\textbf{x}} \geq 1$. This, of course, resembles what happens in empirical observations and here we modify the previous definitions of sRSA and SAR by stipulating that a species can be observed only if it occurs with at least one individual. So, when an area of radius $ R_0 $ harbours $ S(R_0) $ species in total, at smaller radii we define the sRSA as 

\begin{equation}
\texttt{sRSA}(R)=
\frac{ q(N|R)}{\int_1^\infty q(M|R_0) d M}\quad,
\label{def:sRSAemp}
\end{equation}
where $  q(N|R) $ is the distribution that we have obtained in eq.(\ref{eq:sSAD}). From this expression we can derive the SAR, which accounts for the number of species that are found within a certain area as a function of its radius. This is defined as
\begin{equation}
\texttt{SAR}(R)=
S(R_0) \frac{ \int_1^\infty q(N|R) d N}{\int_1^\infty q(M|R_0) d M}\quad.
\label{def:SARemp}
\end{equation}
We have benchmarked the results for the sRSA and SAR obtained from the numerical scheme against the analytical formul\ae \ in figs.(\ref{fig:SAD}) and (\ref{fig:SAR}).

\section{Macroecological patterns of Pasoh and Barro Colorado Island forests}

We considered two datasets from well-known forest stands: one set is from the Barro Colorado Island (BCI) in Panama and the other one from the Pasoh Forest Reserve in Malaysia. Both cover an area of 50 hectars and were comprehensively surveyed, containing high but greatly different numbers of vascular plant species. Species identity, geographical location  and diameter at breast height (DBH) were recorded for each tree living within the plot. We used such datasets of plant species to test model predictions against empirical patterns.
 
\begin{figure}
    \centering
    \begin{subfigure}[t]{.45\textwidth}
        \centering
        \includegraphics[width=\linewidth]{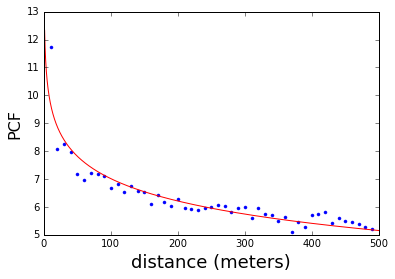} 
    \end{subfigure}
    \hfill
    \begin{subfigure}[t]{.45\textwidth}
        \centering
        \includegraphics[width=\linewidth]{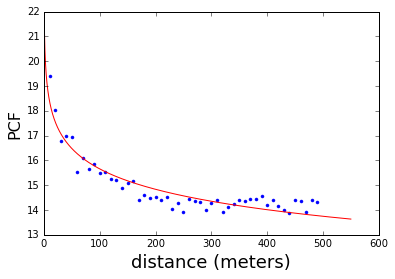} 
          \end{subfigure}

    \vspace{1cm}
    \begin{subfigure}[t]{0.45\textwidth}
           \centering
           \includegraphics[width=\linewidth]{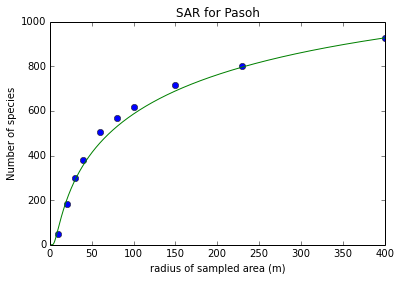} 
       \end{subfigure}
       \hfill
       \begin{subfigure}[t]{0.45\textwidth}
           \centering
           \includegraphics[width=\linewidth]{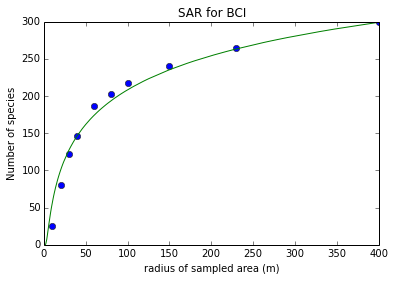} 
             \end{subfigure}
       
        \vspace{1cm}
           \begin{subfigure}[t]{0.45\textwidth}
                  \centering
                  \includegraphics[width=\linewidth]{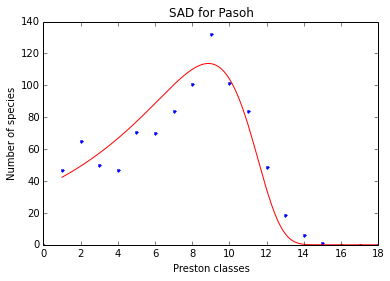} 
              \end{subfigure}
              \hfill
              \begin{subfigure}[t]{0.45\textwidth}
                  \centering
                  \includegraphics[width=\linewidth]{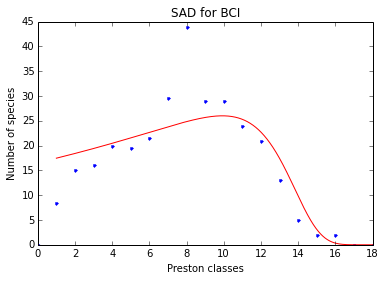} 
                \end{subfigure}
    \caption{The PCF, SAR and Species Abundance Distribution (SAD) for Pasoh (left column) and BCI (right column) tropical forests for trees that are larger than 10cm in stem diameter at breast height. The first panel in each column shows the PCF from which we best-fitted the parameters $ \hat{\lambda} $ and $ \hat{\rho} $ (empirical data showed with blue dots). We found the ratio $\hat{\lambda}/\hat{\rho} \sim 0.33$ for Pasoh, and $\hat{\lambda}/\hat{\rho} \sim 0.35$ for BCI. The second panel depicts the SAR: blue dots are empirical data, green line is the predicted distribution by using the best-fitted parameters from the previous PCF, $ \mean{n} $ and formul\ae \ in eqs.(\ref{def:SARemp}), (\ref{eq:sSAD}) and (\ref{eq:albet}). The third panel shows the SAD (this is defined as the sRSA times the total number of species in the region) for the whole area. The blue dots are empirical data, whereas the red solid line was obtained by using the best-fitted parameters from the previous PCF, $ \mean{n} $ and formul\ae \ in eqs.(\ref{def:sRSAemp}), (\ref{eq:sSAD}) and (\ref{eq:albet}). Preston classes are customarily used in ecological studies and are similar to a $ \log_2 $-binning, although not exactly equivalent. Preston's binning method is described in Volkov et al. (2003).}
    \label{tab:Pasoh_BCI}
\end{figure}

We first coarse-grained the two systems by superimposing a grid mesh of 10m size and counted the number of individuals of each species within every sub-area. Then we looked at each pair of sites located at $\textbf{x}$, $\textbf{y}$ and calculated the empirical PCF with the following formula  
\begin{equation}
g_{\textbf{x},\textbf{y}}=\frac{\frac{1}{S} \sum_{\mu=1}^S n_{\textbf{x}}^{(\mu)} n_{\textbf{y}}^{(\mu)}}{(\frac{1}{S} \sum_{\mu=1}^S n_{\textbf{x}}^{(\mu)})(\frac{1}{S} \sum_{\mu=1}^S n_{\textbf{x}}^{(\mu)})}
\label{eq:correlazione_siti}
\end{equation}
where $n_{\textbf{x}}^{(\mu)}$ is the number of individuals of species $ \mu $ within the site located at $\textbf{x}$ and $S$ is the total number of species in the whole region. Then we calculated the parameters $\hat{\lambda}$ and $\hat{\rho}$ by best-fitting the data to the analytical formula in eq.(\ref{eq:PCF}). Finally, from the empirical data we estimated $ \mean{n}=N_0/(S_0 A_0) $ in both forests, where $ N_0 $ is the total number of individuals, $ S_0 $ is the total number of species in the whole area, $ A_0 $, of the forest plot. We found the ratio $\hat{\lambda}/\hat{\rho} \sim 0.33$ for Pasoh, and $\hat{\lambda}/\hat{\rho} \sim 0.35$ for BCI. These parameters are sufficient to predict the behaviour of the analytical SAR and sRSA with no further best-fit, and such predictions can therefore be compared to the empirical distributions for the two datasets. The agreement with empirical data is good as shown in fig.\ref{tab:Pasoh_BCI}.

\section{Conclusions}
We have introduced a phenomenological stochastic model, defined on a $ d $-dim lattice, from which one can derive analytical approximations of important macro-ecological pattenrs, such as the PCF, the SAR and the sRSA. We devised an efficient numerical integration scheme, which confirms the goodness of the analytical derivations. Also, all the empirical patterns obtained from two canopy forests, the BCI and Pasoh plots, show a good agreement with the formul\ae \ derived from the model, using three free parameters only. The framework is able to explain and link empirical macroecological patterns in a theoretically consistent way. Intriguingly, it suggests that many species-rich ecosystems may possibly be close to a critical point, in which slow and large fluctuations are correlated on large spatial scales. The theoretical setting calls for more refined spatial formulations and better articulated ecological mechanisms, which can provide more realism to the predictions as well as bridge the gap between individual behaviour and emergent macroscale patterns.

\section{Acknowledgements}
The authors would like to thank the Isaac Newton Institute for Mathematical Sciences, Cambridge, for support and hospitality during the programme ``Stochastic Dynamical Systems in Biology: Numerical Methods and Applications'' where work on this paper was undertaken. This work was supported by EPSRC grant n\textdegree \ EP/K032208/1. We are also grateful to the FRIM Pasoh Research Committee (M.N.M. Yusoff, R. Kassim) and the Center for Tropical Research Science (R. Condit, S. Hubbell, R. Foster) for providing the empirical data of the Pasoh and BCI forests, respectively. SA is in debt with Prof. A. Maritan for insightful discussions.

\bibliographystyle{MYnaturemag}
\bibliography{NT_rmp-1}

\end{document}